\documentstyle[prl,aps,twocolumn,psfig]{revtex}
\begin{document}
\draft
\title{EQUATION OF STATE OF THE HARD-SPHERE CRYSTAL}

\author{C. Rasc\'{o}n$^{1,2}$, L. Mederos$^2$, and G. Navascu\'{e}s$^1$}
\address{$^1$Departamento de F\'{\i}sica Te\'{o}rica de la Materia
Condensada, Universidad Aut\'{o}noma, Cantoblanco, Madrid E-28049,Spain}
\address{$^2$Instituto de Ciencia de Materiales (Consejo Superior de
Investigaciones Cient\'{\i}ficas), Cantoblanco, Madrid E-28049, Spain}
\date{\today}
\maketitle

\begin{abstract}
A new approach to the averaged two-particle distribution function of a
crystalline phase is presented. It includes an indirect check of the merit of
the Gaussian approximation for the local density and a new way to inferring
values of the thermodynamic variables from simulation data. The equation of
state and the compressibility of the hard-sphere FCC crystal is computed from
Tarazona free energy density functional [Phys. Rev. A {\bf 31}, 2672 (1985)].
They are in excellent agreement with simulation results over the physical
range of densities up to almost close packing. We also include the comparison
with the results obtained by two other functional approaches which are also
excellent.
\end{abstract}


\section{INTRODUCTION}

     The density-functional theory applied to non-uniform classical fluids
has been able to depict a wide range of physical properties of simple solid
systems. Initially, it was intended for developing a theoretical approach
that was able to describe both fluid and crystalline phases consistently.
Then, it was extended to other crystal properties like elasticity and phonon
dispersion and to more complex systems such as surfaces and liquid crystals
(see reference \cite{Henderson} for a recent review on this matter).

     It is well known the crucial role of hard-spheres (HS) as the usual
reference system of more realistic systems that include attractive
interactions. Accordingly, a considerable effort has been done to develop
free energy functionals describing non-uniform HS systems and, at present,
there exist quite good functional approaches for these systems
\cite{Henderson}. Recently, Denton {\it et al.} \cite{Denton} and Tejero {\it
et al.} \cite{Tejero} have analyzed the equation of state of a HS crystal
obtained from the modified weighted-density approximation (MWDA) \cite{DA}
and from the generalized effective liquid approximation (GELA) \cite{GELA}
respectively. Their analysis include densities well inside the stable solid
phase. This is specially relevant in connection with the solid-solid
transition recently reported by Bolhius and Frenkel \cite{Frenkel} in systems
of HS with a short ranged attractive interaction (we have learnt that Stell
and collaborators had already predicted these kind of transitions in the
seventies \cite{Stell}). It happens that, for sufficiently short-ranged
attractions, this transition occurs at very high densities, near close
packing. To describe this kind of phenomena, the theoretical approach for the
reference HS system should give a reasonable equation of state over all the
physical density range, even proximal to close packing. Thus, in this paper,
we evaluate the equation of state of a face-centered-cubic (FCC) HS crystal
at densities up to almost close packing using the Tarazona
functional approach \cite{T1,T2}. What it is more important is to compare the
functional predictions with simulations results. To do it, we develop a new
method to obtain thermodynamic information from simulation data of
$\tilde{g}(r)$, the average of the two-particle distribution function. This
method also allows us to infer the ideal and the excess contributions (as
defined in section II) to the equation of state. It also corroborates the
merit of the  Gaussian description of the one-particle distribution function.
Finally, but no less important, the method suggest an interesting discussion
on the averaged correlation between the particles beyond nearest-neighbours.
The agreement of the Tarazona functional predictions with simulation data is
excellent up to almost close packing. The same can be said for the
predictions from MWDA and GELA at least up to the densities reported. To test
this accordance, we compute also the compressibility of the HS crystal and
compare it with that of simulation, finding an excellent agreement. Moreover,
we also work out the ideal and excess contributions to the pressure and the
compressibility and compare each one with those corresponding to simulation
data. Again, the accordance is quite good. Note that the evaluation of the
equation of state by MWDA have some problems at high densities as pointed out
by Tejero {\it et al.} \cite{Tejero}.

     In the next section, we briefly resume the Tarazona functional (TF)
together with MWDA and GELA. In section III we present the mentioned
discussion of $\tilde{g}(r)$. The results and a discussion of them are
presented in section IV. The conclusions are exposed in section V.

\section{FUNCTIONAL APPROXIMATIONS}

     Density-functional theories \cite{Henderson} are based on a variational
principle \cite{Mermin} which allows to propose approximations to the
Helmholtz free energy. The variational principle establishes that for a given
interacting potential, fixed external potential and mean density, the
Helmholtz free energy $F[\rho({\bf r})]$, as a functional of the
one-particle density $\rho({\bf r})$, has a minimum value at the equilibrium
density. The free energy functional is usually written as: 

\begin{equation}
F[\rho({\bf r})]=F_{id}[\rho({\bf r})]+F_{ex}[\rho({\bf r})],
\label{F}
\end{equation}
where $F_{id}$ is the ideal contribution to the Helmholtz free energy of
a non-uniform system. Its functional form is exactly given by:

\begin{equation}
\beta F_{id}[\rho({\bf r})]=
\int d{\bf r} \rho({\bf r}) (ln(\Lambda^3\rho({\bf r})) -1),
\label{Fid}
\end{equation}
where $\Lambda$ is the thermal de Broglie wavelength and $\beta=1/k_BT$. The
second term of (\ref{F}), $F_{ex}$, called free energy excess, arises from
the interacting potential between particles. Several accurate approximations
have been proposed for the free energy excess. These are based on a mapping
of the thermodynamic properties of the non-uniform systems onto those of a
uniform fluid at some effective density (weighted-density). Tarazona
\cite{T1,T2} proposes the following expression for the free energy
excess of the HS solid:

\begin{equation}
F_{ex}[\rho({\bf r})] = \int d{\bf r} \rho({\bf r})\Delta\Psi_{ex}
(\bar{\rho}({\bf r})),
\label{TFex}
\end{equation}
where $\Delta\Psi_{ex}({\bf \rho})$ is the free energy excess per particle of
the uniform system at density $\rho$ and the weighted density, $\bar{\rho}$,
is given by

\begin{equation}
\bar{\rho}({\bf r}) = \int d{\bf r}' \rho({\bf r}')  w(|{\bf r}-{\bf
r}'|;\bar{\rho}({\bf r})).
\label{Tbarra}
\end{equation}
The function $w(r)$, in the integral equation that defines $\bar{\rho}$, is
specified by requiring that
the direct correlation function $c(r)$ obtained from the free energy
functional matches that of the HS liquid in the uniform limit.
In the MWDA, Denton and Ashcroft
use the same mapping idea. However, they  propose a global map of the
free energy excess onto the free energy of a unique uniform fluid
\cite{DA,Denton}:

\begin{equation}
F_{ex}[\rho({\bf r})] = N \Delta\Psi_{ex}(\bar{\rho}),
\label{DFex}
\end{equation}
where the weighted density $\bar{\rho}$ is given by 

\begin{equation} 
\bar{\rho}={1 \over N} \int d{\bf r} \rho({\bf r}) \int d{\bf r}' \rho({\bf
r}')  w(|{\bf r}-{\bf r}'|;\bar{\rho}).
\label{Dbarra}
\end{equation}
In the GELA, Lustko and Baus use again a global map but the weighted density
(or effective density), $\bar{\rho}$, is given by the structural mapping 
\cite{GELA}:

\begin{displaymath}
\int d{\bf r} \rho({\bf r}) \int d{\bf r}' \rho({\bf r}')  c(|{\bf r}-{\bf
r}'|;\bar{\rho})=
\end{displaymath}
\begin{equation}
=\int d{\bf r} \rho({\bf r}) \int d{\bf r}' \rho({\bf r}')  c({\bf r},{\bf
r}';{\rho({\bf r}})),
\label{Bbarra}
\end{equation}
where $c$ is the direct correlation function.

An important advantage of the two latter approaches is that they require less
computational effort. However, only the TF is a true functional approach in
the sense that it is not restricted to macroscopically homogeneous systems, as
the present case of the HS crystal. For more technical details, the reader is
referred to references \cite{T2}, \cite{DA} and \cite{GELA} in relation to
TF, MWDA  and GELA respectively and to reference \cite{Henderson} for an
extensive discussion of these and other functional approximations.

     The one-particle density, $\rho({\bf r})$, is usually assumed to be
described by a sum of normalized Gaussians:

\begin{equation}
\rho({\bf r}) ={{({\alpha \over \pi})}^{3 \over 2}} \sum_{\bf R}
e^{-\alpha({\bf r}-{\bf R})^2},
\label{Gaus}
\end{equation}
where ${\bf R}$ is the vector position of the crystal lattice and $\alpha$ is
the Gaussian width parameter. With this density parametrization, the
variational principle reduces to finding the value of $\alpha$ that minimizes
the free energy functional at each mean density. Several attempts trying to
improve the parametrization of the one-particle density have shown the
goodness of the Gaussian one \cite{Haymet}. On the other hand, simulations
have shown that deviations from Gaussian form are only significant at low
densities but only at the tails of the distribution \cite{Wagner}. For those
reasons, the Gaussian parametrization seems to be an excellent description of
the one-particle density of the HS crystal over all range of physical
densities. We shall give another evidence based on MC simulations 
\cite{Weis,Weis2,Jackson,Ree86,Ree91}.

     The equation of state for the HS solid is obtained as follows. After the
minimization process we obtain the free energy per particle
$f(\rho)=f(\rho,\alpha(\rho))$ at each mean density $\rho$. Notice that
$\alpha$ is, after the minimization, a function of the mean density $\rho$.
Then, at fixed temperature, the pressure is given by:

\begin{equation}
{{\beta P} \over \rho} = \beta \rho {{\partial f(\rho)} \over
{\partial \rho}}.
\label{p}
\end{equation}
Because the standard division of the Helmholtz free energy (\ref{F}) into the
ideal contribution and the excess one, it follows the equivalent for the free
energy per particle, namely $f=f_{id}+f_{ex}$. Therefore, the equation of
state (\ref{p}) can be formally split into two terms:

\begin{equation}
{{\beta P_{id}} \over \rho} = \beta \rho {{\partial f_{id}(\rho)} \over
{\partial \rho}}
\label{pid}
\end{equation}
\begin{equation}
{{\beta P_{ex}} \over \rho} = \beta \rho {{\partial f_{ex}(\rho)} \over
{\partial \rho}},
\label{pex}
\end{equation}
where $P=P_{id}+P_{ex}$. Following Denton {\it et al.}, we call $P_{id}$ and
$P_{ex}$ 'ideal-gas pressure' and 'excess pressure' respectively. Notice that
$P_{id}$ is not the usual ideal gas pressure, i.e., that which gives the
ideal compressibility factor. In addition to the different functional
approach used by Denton {\it et al.} for the Helmholtz free energy excess,
these authors approximate the ideal free energy per particle by:

\begin{equation}
\beta f_{id}(\alpha)={3 \over 2} ln({\alpha \over \pi})+3 ln(\Lambda)-{5
\over 2},
\label{fid}
\end{equation}
which is exact in the limit of non overlapping Gaussians. It is a very good
approximation over all density range of the HS solid. However, we use the
exact functional expression (\ref{Fid}) to obtain the TF ideal contribution.

     An important and direct test for the equation of state is its capacity
to predict the isothermal compressibility of the HS crystal. This is given
by:

\begin{equation}
\rho k_BT\chi_T =\left[{{\partial \beta P(\rho) } \over {\partial
\rho}}\right]^{-1} ,
\label{X}
\end{equation}
which, extending the above formal division into an ideal part and an excess
part, can be written as:

\begin{equation}
\rho k_BT\chi_{id} =\left[{{\partial \beta P_{id}(\rho) } \over{\partial
\rho}}\right]^{-1} ,
\label{Xid}
\end{equation}
\begin{equation}
\rho k_BT\chi_{ex} =\left[{{\partial \beta P_{ex}(\rho) } \over{\partial
\rho}}\right]^{-1} ,
\label{Xex}
\end{equation}
with

\begin{equation}
      {1 \over \chi} = {1 \over \chi_{id}} + {1 \over \chi_{ex}} .
\label{XXX}
\end{equation}

\section{THE AVERAGE OF THE TWO PARTICLE DISTRIBUTION FUNCTION}

 The averaged two-particle distribution, $\tilde{g}(r)$, is defined as:

\begin{equation}
{\tilde{g}(r_{12})} ={{1} \over {4 \pi V \rho^2}} \int d\Omega_{12} \int
d{\bf r}_2  \rho^{(2)}({\bf r}_1,{\bf r}_2),
\label{gtilde}
\end{equation}
where $V$ is the volume,
$\rho^{(2)}({\bf r}_1,{\bf r}_2)$ the two-particle density function
and $d\Omega_{12}$ the differential solid angle aperture
around ${\bf r}_{12}$.
In the uniform limit Eqn.(\ref{gtilde}) reduces to the well known radial
distribution function. MC results for this function were parametrized
originally by Weis \cite{Weis} with the following analytical expression:

\begin{equation}
\tilde{g} (r) = 0,\;\;\;0\leq r \leq 1
\label{gtildeREE0}
\end{equation}
and
\begin{equation}
\tilde{g} (r) =  \sum_{i\geq 1} \tilde{g}^{(i)}(r),\;\;\;r\geq 1,
\label{gtildeREE1}
\end{equation}
with
\begin{equation}
\tilde{g}^{(1)}(r) = {A \over r} e^{-[W_1(r-r_1)]^2-[W_2(r-r_1)]^4}
\label{gtildeREE2}
\end{equation}
and
\begin{equation}
\tilde{g}^{(i)}(r) = {1 \over {4 \pi \rho}}({W^2 \over \pi})^{1\over
2}\:  n_i\, {e^{-[W(r-R_i)]^2}\over R_i r},\;\;\;i\geq 2,
\label{gtildeREE3}
\end{equation}
where $n_i$ is the number of lattice sites at distance $R_i$ and A is
determined by the virial theorem. The parameters $r_1$, $W_1$, $W_2$ and $W$
are elaborated analytic functions of the density that give a good fitting to
results of MC simulation (distances in all above equations are given in HS
diameter units). These functions were refined successively
\cite{Weis2,Jackson,Ree86}. The parameters provided by Choi {\it et al.}
\cite{Ree91} overcome those of previous authors giving an accurate
description of the MC results. The maximum root mean square deviation of
$\tilde{g}(r)$ from MC data computed over the distance range (r up to 3.3) is
0.17 at the highest density. It quickly decreases to less than 0.06 at
packing fractions lower than $\eta=0.65$.

     Here, the interesting point is the functional form used to fit the MC
data. The function $\tilde{g}(r)$ is written as a sum of peaks corresponding
to successive shells of neighbours. The first peak, Eq.(\ref{gtildeREE2}),
has a characteristic form but all the rest, Eq.(\ref{gtildeREE3}), have
exactly the same functional form. These later peaks only differ from each
other on the distances where they are located, $R_i$, and on their
normalizations which correspond to the number of neighbours at distance $R_i$
(according to a lattice without vacancies).

     The crucial point is to realize that if the spherical average of
$\rho^{(2)}({\bf r},{\bf r}')$, which defines $\tilde{g}(r)$, is done on the
product of two sums of Gaussians, $\rho({\bf r})\rho({\bf r}')$, we obtain
precisely all the peaks of $\tilde{g}(r)$ except the first one. In effect,
from the definition of this average:

\begin{equation}
{\tilde{g}_0(r_{12})} ={{1} \over {4 \pi V \rho^2}} \int d\Omega_{12} \int
d{\bf r}_2  \rho({\bf r}_1)\rho({\bf r}_2),
\label{ggaus}
\end{equation}
it is straightforward to obtain

\begin{equation}
{\tilde{g}_0(r)} =  \sum_{i\geq 0} \tilde{g}^{(i)}_0(r),\;\;\;r\geq 0
\label{ggaussuma}
\end{equation}
with
\begin{equation}
\tilde{g}^{(0)}_0(r) = 
{1\over{4 \pi \rho}}({\alpha\over{2 \pi}})^{1\over2}2\alpha e^{-{\alpha \over
2}r^2},
\label{ggaus0}
\end{equation}
and
\begin{equation}
\tilde{g}^{(i)}_0(r) = 
{1 \over {4 \pi \rho}}({\alpha \over {2 \pi}})^{1\over 2} n_i
{e^{-{\alpha \over 2}(r-R_i)^2}+e^{-{\alpha \over 2}(r+R_i)^2}\over R_i r}
\label{ggausia}
\end{equation}
\begin{equation}
={1 \over {4 \pi \rho}}({\alpha \over {2 \pi}})^{1\over 2}
 n_i {e^{-{\alpha \over 2}(r-R_i)^2}\over R_i r},
\;\;\;i\geq 1.
\label{ggausib}
\end{equation}
We have dropped the second exponential in the expression (\ref{ggausia}) as
its contribution to each peak is, at most, 20 orders of magnitude smaller
than the first exponential (for the current values of the Gaussinan parameter
$\alpha$). The first peak, Eq.(\ref{ggaus0}), of $\tilde{g}_0(r)$ does not
appear in $\tilde{g}(r)$ because the exclusion of the self-interaction. The
second one, Eq.(\ref{ggausib}) for $i=1$, differs from the first one,
Eq.(\ref{gtildeREE2}), of $\tilde{g}(r)$. However, identifying $\alpha \over
2$ with $W^2$, all the remaining peaks are exactly the same both in
$\tilde{g}(r)$ and in $\tilde{g}_0(r)$. 

     The identification of peaks in $\tilde{g}(r)$ with those in
$\tilde{g}_0(r)$ has several interesting consequences. The immediately
obvious one is that the whole two-body correlation between particles beyond
nearest-neighbours is already included in the product $\rho({\bf r})\rho({\bf
r}')$. This point has been suggested and extensively discussed by two of us
in relation with the perturbation weighted density approximation (PWDA) for
simple systems with attractive interaction potentials \cite{PWDA1,PWDA2} and
it is confirmed in the present discussion. We have proposed that most of the
correlation of the two particle distribution function, 

\begin{equation}
\rho^{(2)}({\bf r},{\bf r}')\equiv\rho({\bf r})\rho({\bf r}')g({\bf r},{\bf
r}'),
\label{roro}
\end{equation}
is already described by the mentioned product. Thus, $g({\bf r},{\bf r}')$
could be basically approximated by a step function to exclude the
self-interaction. Instead, we went further and mapped $g({\bf r},{\bf r}')$
into a homogeneous fluid at an very low effective density determined by the
compressibility equation. This proved to be an excellent criterium to
determine the perturbation contribution to the free energy \cite{PWDA2}.

     On the other hand, the identification of peaks also confirms the
goodness of the Gaussian parametrization for the one-particle density.
Moreover, it suggests that from the theoretical minimization process of the
free energy one can obtain information on the crystal average distribution
function $\tilde{g}(r)$. (A work along this line is in progress).  Here, we
explore the other way: from MC data (the parameter W in (\ref{gtildeREE3}))
we obtain the corresponding Gaussian width parameter $\alpha$ of the 'MC
Gaussian one-particle density'. Then, we are able to compute the MC ideal
contribution to the free energy, $F_{id}$, throughout the exact functional
form Eq.(\ref{Fid}). The MC free energy excess is now immediately obtained
and so are the ideal and the excess contributions of the pressure and the
compressibility. Notice that the excellent fitting of Choi {\it et al.}
allows us to evaluate analytically all these thermodynamics properties over
the density range up to almost close packing. In figure 1, for $\alpha$, and
figure 4, for $P_{id}$ and $P_{ex}$, we show some of these results in
comparison with those inferred by Denton {\it et al.} \cite{Denton} from
Young and Alder \cite{YA} simulation data. The fair agreement is a sign of
the consistency of the different MC data. In any case, the treatment of MC
data we have followed is much more powerful as it gives a continuous
expression of the thermodynamic variables,  as a function of density, up to
almost close packing.

\section{RESULTS AND DISCUSSION}

     Figure 1 shows the logarithm of the Gaussian parameter $\alpha$ inferred
by us from Choi {\it et al.} MC simulations \cite{Ree91} and  by Denton {\it
et al.} \cite{Denton} from Young and Alder MC simulations  \cite{YA}. The
predictions of TF, MWDA and GELA are also displayed. (MWDA data in all
figures are those obtained by Tejero {\it et al.} which we assume free of
possible convergence problems). TF predicts a systematic overestimation of
the Gaussian width parameter $\alpha$ but always in fair agreement with
simulation results. GELA predictions are better at low densities as it was
already known. At high densities both, TF and GELA, predict close values for
$\alpha$. The MWDA gives the best predictions except at the lowest densities.
     Figure 2 shows the equation of state predicted by TF, MWDA and GELA
together with simulation results. At low densities (see also Figure 3), the
GELA pressure is slightly above the simulation pressure while TF and MWDA
pressures have lower values. All of them improve their agreement with
simulation as the pressure increases. At the highest density where we have
performed calculations, the TF pressure differs from the simulation one
in less
than $0.3\%$. In Figure 4, we show the two contributions to
the pressure, ideal and excess, obtained from TF, MWDA and GELA 
in comparison with those
inferred from simulation data. TF and GELA show an excellent agreement with
simulation results. The MWDA results are also quite good.
      Notice that the good theoretical predictions for the $\alpha$ 
parameter (Figure 1) will necessarily mean comparable good predictions 
for the ideal pressure (Figure 4) because equation (\ref{Fid}) is exact.
However, due to the functional approximations proposed for the excess
free energy (or for the total free energy) the theoretical excess
pressure (or the total pressure) must be also compared with simulation
data. This is even more important at high densities where a significant
error in the excess contribution would not have appreciable effects  on
the total pressure.
      Figure 5 shows the compressibility predicted by TF together with that
obtained from simulation. We have used the standard cubic spline treatment of
the pressure data to obtain the compressibility via equation (\ref{X}). Once
more, the agreement is quite good.
We do not have accurate enough data of GELA and MWDA 
to obtain the compressibility properly. It exhibits unphysical oscillations
which, most probably, are due to round off effects of pressure data. For
completeness, we present in inset of figure 5 the inverse of the ideal and
excess compressibilities. The agreement with simulation is good.

      The splitting of the pressure into those parts is quite convenient for
all the above discussion. However, we are inclined to think that there is no 
physical meaning in the ideal and the excess pressures, such as they are
defined by Eq.(\ref{pid}) and (\ref{pex}).

\section{CONCLUSIONS}

     In this paper, we have presented two different contributions. One
concerns to the method used to handle MC data of the $\tilde{g}(r)$ and the
other concerns to the accuracy of the equation of state computed with
functional approaches.

     We have proposed a new method, using MC data of $\tilde{g}(r)$, to
determine the rms displacement of a particle from its equilibrium lattice
site (i.e. the parameter $\alpha$ if $\rho({\bf r})$ is approximated by a sum
of Gaussians). This should work perfectly for peaks of $\tilde{g}(r)$ at
distances where the two-particle distribution function is already given by
the product $\rho({\bf r})\rho({\bf r}')$. In practice, it is possible to
deduce already the rms displacement (or parameter $\alpha$) from the second
peak of $\tilde{g}(r)$. This procedure should be consistent with a direct
determination of the rms displacement from MC configurations. Assuming the
Gaussian parametrization for $\rho({\bf r})$,  an accurate fitting of
$\tilde{g}(r)$ allows us to determine the parameter $\alpha$ as a function of
the density. From this, the ideal pressure is straightforwardly obtained via
the exact expression (\ref{Fid}). From the total and the ideal pressures, the
excess pressure follows immediately.

      As a consequence of the analytical fitting of MC data for
$\tilde{g}(r)$, it can be deduced that beyond the nearest-neighbours the
two-particle correlation function is practically given by the product
$\rho({\bf r})\rho({\bf r}')$, i.e., $g({\bf r},{\bf r}')\approx 1$. We are not
aware that this quite interesting and important result was previously
mentioned by other authors.

     The other aim of this paper has been to show the excellent behaviour of
the equation of state of the FCC HS solid computed with functional
approaches. They agree with simulation results up to almost close packing.
This remarkable behaviour is extended to the contributions to the pressure,
namely the ideal and the excess pressures, and to the compressibility and
their ideal and excess parts. They all are in notable agreement with
simulation up to almost close packing.

      Finally, we want to remark that functional approaches provide a
reliable reference HS system for perturbation theories, especially at high
densities. This makes them particularly suitable for describing the
solid-solid isostructural transition of simple systems with extremely short
ranged attractive potentials \cite{Frenkel}. We have applied the PWDA
\cite{PWDA1,PWDA2} mentioned above to an attractive square well \cite{pozo}
and to HS plus Yukawa attractive tail \cite{Yukawa}. Precisely, the PWDA is
based, among other things, on two of the points explored here: the goodness
of the TF, even at high densities, for describing HS and the accuracy of the
product $\rho({\bf r})\rho({\bf r}')$ for describing the correlation between
particles beyond nearest neighbours. 
Note that after Tejero {\it et al.} discussion on
the numerical problems at high densities of the MWDA, results based on this
functional approach must be seem with some caution \cite{Lik}.

\acknowledgments

This work was supported by the Direcci\'{o}n General de Investigaci\'{o}n
Cient\'{\i}fica y T\'{e}cnica of Spain, under Grant $PB94-0005-C02$.

\begin{figure}
\caption{Logarithm of the  Gaussian width parameter $\alpha$ {\it vs.}
packing fraction $\eta$. 
Solid curve inferred, in this paper, from simulation data of Choi {\it et al.}.
Solid squares inferred by Denton {\it et al.} from
simulation data of Young and Alder. 
The dashed curve is the TF prediction. Solid triangles and open triangles
are the results from GELA and MWDA respectively
obtained by Tejero {\it et al.}.}
\end{figure}

\begin{figure}
\caption{Logarithm of the equation of state $P/\rho k_BT $
{\it vs.} packing fraction $\eta$.
Curves and symbols as in figure 1.}
\end{figure}

\begin{figure}
\caption{Equation of state $P/\rho k_BT $ {\it vs.} packing
fraction $\eta$ at low densities.
Curves and symbols as in figure 1.}
\end{figure}

\begin{figure}
\caption{Equation of state $P/\rho k_BT $ {\it vs.} packing fraction $\eta$.
Curves and symbols as in figure 1.}
\end{figure}

\begin{figure}
\caption{$\rho k_BT\chi_T$ {\it vs.} packing fraction $\eta$.
Inset: $1/ \rho k_BT\chi_T$ {\it vs.} packing fraction $\eta$. 
Curves as in figure 1.}
\end{figure}

\end{document}